\documentclass[%
 reprint,
 amsmath,amssymb,
 aps,
 prl,
 longbibliography,
 lengthcheck,%
]{revtex4-1}

\usepackage{graphicx}
\usepackage{dcolumn}
\usepackage{bm}
\usepackage{hyperref}

\begin{document}

\preprint{APS/123-QED}

\title{Waves of spin-current in magnetized dielectrics
}

\author{P. A. Andreev}%
\email{andreevpa@physics.msu.ru}
\affiliation{
 Department of General Physics,
 Faculty of Physics, Moscow State
University, Moscow, Russian Federation.}
\author{L. S. Kuz'menkov}%
 \email{lsk@phys.msu.ru}
\affiliation{%
 Department of Theoretical Physics, Physics Faculty, Moscow State
University, Moscow, Russian Federation.}%




\date{\today}

\begin{abstract}
Spin-current is an important physical quantity in present day spintronics and it might be very usefull in the physics of quantum plasma of spinning particles. Thus it is important to have an equation of the spin-current evolution. This equation naturally appears as a part of a set of the quantum hydrodynamics (QHD) equations. Consequently, we present the set of the QHD equations derived from the many-particle microscopic Schrodinger equation, which consists of the continuity equation, the Euler equation, the Bloch equation and equation of the spin-current evolution. We use these equations to study dispersion of the collective excitations in the three dimensional samples of the magnetized dielectrics. We show that dynamics of the spin-current leads to formation of new type of the collective excitations in the magnetized dielectrics, which we called spin-current waves.
\end{abstract}

\maketitle


\section{\label{sec:level1}I. Introduction}

The spin-current is a very important characteristic for various physical systems. For instance it is useful for
description of such processes as spin injection ~\cite{Takahashi JMMM
04} and other process, where spin transport is involved. Most of them are
accumulated in a grate field of physics: spintronics ~\cite{Zutic
RMP 04},~\cite{Sinova Nat Mat 12}. There are different methods of the
spin-current generation, such as spin pumping by sound waves
~\cite{Uchida JAP 12}, optical spin injection ~\cite{Pezzoli PRL
12}, and at using of junctions as Polarizers ~\cite{Rashidian IJTP
12}, spin flip in the result of electron interaction with
electromagnetic wave ~\cite{Hammond APL 12}. Spin-polarized
currents are used in spin lasers development ~\cite{Lee PRB 12}
and in spin-diode structures ~\cite{Li JAP 12}. It is interesting
to admit that in some devises the spin-current exhibits the sine
wave-like behavior ~\cite{Zhang P B 12}. New modification of spin-field effect transistors have been suggested (see for example
~\cite{Ma JAP 12}, ~\cite{Bagraev P C 10}), one of the spintronic
devices utilize the electron's spin properties in addition to its
charge properties. Effects in junction play key role in spintronic
devises. The spin-dependent Peltier effect was observed
experimentally
~\cite{Flipse Nat Nano 12}. It is based on the ability of the
spin-up and spin-down channels to transport heat independently. The use of
graphene has been involved in this field either ~\cite{Pesin Nat Mat 12}, ~\cite{Vera-Marun Nat
phys 12}. Processes of spin
transport and relaxation in graphene have also been considered
~\cite{Han JMMM 12}. It has been expected that silicon spintronics has
potential to change information technology, this possibility
discussed in Ref. ~\cite{Jansen Nat Mat 12}.

We have described a number of examples there the spin-current plays important role. Thus, it is necessary to have analytical definition of the spin-current and equation of the spin-current evolution. In Refs. ~\cite{An Sci Rep 12}-~\cite{Shi PRL 06} authors discussed the definition of the spin-current in terms of the one-particle wave function describing the quantum state of the spinning particle. Knowledge of the wave function allows to calculate the spin-current and make conclusions on the system behavior. We keep the spin-current as an independent macroscopic variable defined via the many-particle wave function governs system behavior. We use the spin-current along with the particle concentration, the velocity field, and the magnetization. We consider the spin-current evolution due to interparticle interaction. For this purpose we derive the spin-current evolution equation as a part of the set of quantum hydrodynamic equations. Total angular momentum current was presented in Ref. ~\cite{An Sci Rep 12} as a generalization of the spin current, which conserve even when the spin-current does not conserve. In our paper we consider the spin-current only, since the spin-current has appeared in the quantum hydrodynamic equations. It is not necessary to have a conserved quantity for the spin-current. We study the spin-current evolution due to interparticle interaction.

It can be expected that the spin-current evolution also gives influence on the spin waves,
which are well-known phenomenon in many different physical
systems, first of all they exist in the ferromagnetics and other
structures with strong magnetization. For example quantum dots
show interesting spin waves behavior ~\cite{Aliev PRB
09}-~\cite{Guslienko PRB 10}. Dynamics of the magnetic moments in the quantum
plasma has also been studied. In Ref. ~\cite{Andreev VestnMSU
2007} existence of the self-consistent spin waves in the magnetized
plasma was shown. Thus new branches of the wave dispersion appear in the
magnetized plasma due to the dynamics of magnetic moment of electrons
and ions. It was demonstrated by means of the quantum
hydrodynamics (QHD) method ~\cite{MaksimovTMP 1999}-~\cite{Andreev
arxiv 11 3}. Later, the generalization of the Vlasov equation
for the plasma of spinning charged particles was used to study the
same problem ~\cite{Brodin PRL 08}. Magnetic moment dynamics also
leads to existence of the effect of resonances interaction of the
neutron beam with the magnetized plasma, which gives new method of wave
generation in the magnetized plasma ~\cite{Andreev IJMP 12}, ~\cite{Andreev AtPhys 08}, ~\cite{Andreev PIERS 2011}.
Method of the QHD has become very popular and powerful method of
studying influence of the magnetic moment dynamics on various
processes in the magnetized plasma ~\cite{Andreev IJMP 12},
~\cite{Maksimov VestnMSU 2000}-~\cite{Shukla RMP 11}.
The method of the QHD can be used for systems of the neutral
particles with the magnetic moment. Such systems are more preferable
to demonstrate results given by the QHD method at description of the
physical effects caused by evolution of the magnetic moments.
The QHD description of the self-consistent spin waves in a system of
neutral particles in the one-, two-, and three-dimensional dielectrics
was described in Ref. ~\cite{Andreev PIERS 2012 Spin}. Usually the
set of QHD equations for the spinning charged particles consists
of the three equations for evolution of the material fields, these are the
continuity equation for the particles concentration, the Euler
equation for the velocity field, and the generalized Bloch
equation for the magnetic moment evolution, and the Maxwell
equations for the electromagnetic field description.

The spin-current and an equation of its evolution are in the
center of attention of this paper. The spin-current for many-particle systems of charged spinning particles
appears in the QHD equations ~\cite{Andreev IJMP 12}, ~\cite{MaksimovTMP 2001}. Usually we need to find approximate relation between the spin-current and other hydrodynamic variables to get closed set of the QHD equations. In this paper we go further, we derive an equation for the spin-current
evolution by means of the QHD method. Recently, an analogous equation was
derived for graphene electrons which have also been obtained by the QHD
method ~\cite{Andreev arxiv 12 01}. The spin-current definition and corresponding equations appears in the QHD in the
semi-relativistic approximation. The spin-current have
been universally defined according to the quantum electrodynamics
in Ref. ~\cite{An Sci Rep 12}.

The first step in the development of the QHD was made after the Schrodinger equation had been suggested. E. Madelung represented the Schrodinger equation for the one charged particle in an external field as a set of two equations \cite{Madelung 26}. These equations are the continuity equation and the Euler equation for the eddy-free motion. Later T. Takabayasi considered the QHD representation of the Pauli equation \cite{Takabayasi PTP 55}.

It is well-known that the one-particle Pauli equation can be represented as the set of the QHD equations. This set of equations is almost equivalent to the first three equations of the chain of the many-particle QHD equations taken in the self-consistent field approximation. The differences are the follows. One-particle equations do not contain the thermal pressure and the thermal contribution in the spin-current. Hence, the one-particle approximation may be used in semiconductors at low temperatures. However, the hydrodynamical equations should be coupled with the set of Maxwell equations.

If we derive the QHD equations from the Pauli equation for the
one-particle in an external field we find three equation of
material field evolution mentioned above. However, if we have deal
with many-particle system the set of the QHD equations contains
some new functions, for example the kinetic pressure $p^{\alpha\beta}$
caused by the thermal motion of particles, and the spin-current
$J^{\alpha\beta}$. Two-particle macroscopic functions also appear
in the terms describing inter-particle interaction. In the
self-consistent field approximation, containing no contribution of the
exchange interaction, the terms describing interaction have same
form as the terms caused by the external fields.

To continue the comparison of the many-particle QHD with the
one-particle one we admit that in the one-particle case
$p^{\alpha\beta}=0$ and $J^{\alpha\beta}=M^{\alpha}v^{\beta}$,
where $M^{\alpha}$ is the density of magnetic moments, and
$v^{\beta}$ is the velocity field. We can also notice that in the
one-particle case the kinetic energy field $\varepsilon$ easily related with
the particle concentration $n$, magnetization $M^{\alpha}$ and
velocity field $v^{\beta}$, so we have
$\varepsilon=mnv^{2}/2$. For the many-particle system we have
that $p^{\alpha\beta}$, $J^{\alpha\beta}$ and $\varepsilon$
are independent material fields additional to $n$, $v^{\alpha}$,
and $M^{\alpha}$. In the many-particle system $J^{\alpha\beta}$ and
$\varepsilon$ are partly connected with $n$, $v^{\alpha}$, and
$M^{\alpha}$. Let us make an example to describe the last statement.
For the energy we have $\varepsilon=mnv^{2}/2+n\epsilon$,
where the last term corresponds to internal energy caused by the
thermal motion of particles \cite{MaksimovTMP 1999}. A set of QHD equations including the
energy evolution equation and the non-zero thermal pressure in the
Euler equation can be derived from a quantum kinetic equation, but
derivation of the quantum kinetic equation needs some additional
assumption to construct a distribution function \cite{Balescu book Eq Stat mech}, \cite{Wigner PR 84}. Thus, to make more detailed study of magnetic moment dynamics we are
going to present the straightforward derivation of equation for the magnetic moment current (or the
spin current) $J^{\alpha\beta}$ evolution from the many-particle Schrodinger and study its influence on the
spin wave dispersion.
During several last
decades a lot of different ways of the quantum kinetic
equation derivation were suggested, but there are a lot of open questions in
this field. Derivation of a kinetic equation via the Wigner
distribution function is the most popular and actively used in
recent publications. Moreover we keep developing the method of
many-particle quantum hydrodynamics which is more direct way of
derivation of equations for collective quantum dynamics \cite{Andreev IJMP 12}, \cite{Andreev Qkin 12}, \cite{Andreev Asenjo 13}.

Waves in systems of neutral and charged spinning particles have been considered by means of many-particle quantum hydrodynamics with no account of the spin-current equation. It has been assumed that the spin-current $J^{\alpha\beta}$ appearing in the magnetic moment evolution equation can be approximately considered as $J^{\alpha\beta}=M^{\alpha}v^{\beta}$. In this paper studying waves in the magnetized dielectrics we consider two kind of equilibrium states. One of them
corresponds to the case when the equilibrium spin current
$J^{\alpha\beta}$ equals to zero, in second case we consider a non
zero equilibrium spin current $J^{\alpha\beta}$, but we suppose
that the equilibrium velocity field equals to zero. Such structure
might be realized by means two currents (flows of neutral
particles) directed in opposite directions and having opposite
equilibrium spin.

We also need to accent the fact that the many-particle QHD method
has been used for the different physical systems, thus the sets of the
QHD equations have been obtained for graphene ~\cite{Andreev arxiv 12
01}, the neutral ultracold quantum gases ~\cite{Andreev arxiv 12
transv dip BEC}, the Bose-Einstein condensate of excitons in graphene
~\cite{Andreev arxiv 12 GEBEC}, along with the physics of plasma
described above.

Presented here results are also important for the physics of
magnetized ultracold quantum gases. Used where models are
equivalent to the first three equations of the QHD, they are the
continuity equation, the Euler equation, and the Bloch equation
~\cite{Machida JPSJ 98}, ~\cite{Ho PRL 98}.

This paper is organized as follows. In Sec. II  we present and
describe the set of the QHD equations derives in the paper. In
Sec. III dispersion of the spin waves is considered, a contribution of
the spin-current in the spin wave properties is studied. In Sec. IV
brief summary of obtained results are presented.

\section{II. The model}

The many-particle QHD equations are derived from the microscopic many-particle Schrodinger equation $\imath\hbar\partial_{t}\Psi(R,t)=\hat{H}\Psi(R,t)$, where $\Psi(R,t)$ is the wave function of $N$ interacting particles. $\Psi(R,t)$ depends on coordinate of all particles. We present it shortly by means $R$, which is $R=(\textbf{r}_{1},\textbf{r}_{2},...,\textbf{r}_{N})$, where $\textbf{r}_{i}$ is coordinate of $i$ th particle. The structure of the QHD equations depends on the explicit form of the Hamiltonian of considered system of particles. We do not described here the method of derivation of the QHD equations, a lot of paper are dedicated to this topic  ~\cite{MaksimovTMP 1999}-~\cite{Andreev arxiv 11 3}. However, to be certain we present the Hamiltonian
$$\hat{H}=\sum_{p}\biggl(\frac{1}{2m_{p}}\textbf{D}^{2}_{p}+e_{p}\varphi^{ext}_{p}-\gamma_{p}\sigma^{\alpha}_{p}B^{\alpha}_{p(ext)}\biggr)$$
\begin{equation}\label{SC ham gen}+\frac{1}{2}\sum_{p,n\neq p}(e_{p}e_{n}G_{pn}-\gamma_{p}\gamma_{n}G^{\alpha\beta}_{pn}\sigma^{\alpha}_{p}\sigma^{\beta}_{n})\end{equation}
used for derivation of equations presented below, where
$D_{p}^{\alpha}=-\imath\hbar\partial_{p}^{\alpha}-e_{p}A_{p,ext}^{\alpha}/c$,
$\varphi_{p,ext}$, $A_{p,ext}^{\alpha}$ are the potentials of the
external electromagnetic field,
$\partial_{p}^{\alpha}=\nabla_{p}^{\alpha}$ is the derivatives on space variables,
and $G_{pn}=1/r_{pn}$ is the Green functions of the Coulomb
interaction, $G^{\alpha\beta}_{pn}=4\pi\delta^{\alpha\beta}\delta(\textbf{r}_{pn})+\partial^{\alpha}_{p}\partial^{\beta}_{p}(1/r_{pn})$ is the Green function of spin-spin interaction, $\gamma_{p}$ is the gyromagnetic ratio, $\sigma^{\alpha}_{p}$ is the Pauli matrix, a commutation relations
for them is
$$[\sigma^{\alpha}_{p},\sigma^{\beta}_{n}]=2\imath\delta_{pn}\varepsilon^{\alpha\beta\gamma}\sigma^{\gamma}_{p},$$
$e_{n}$, $m_{n}$ are the charge and the mass of particle, $\hbar$ is the Planck constant and $c$ is the speed of light. For electrons
$\gamma_{p}$ reads $\gamma_{p}=e_{p}\hbar/(2m_{p}c)$,
$e_{p}=-|e|$.

This Hamiltonian (\ref{SC ham gen}) consists of two parts. The first of them is presented by the first three terms, which describe motion of independent particle in an external electromagnetic field. The first term is the kinetic energy, which contains vector potential via covariant derivative $\textbf{D}_{n}$. Therefore, it contains action of an external magnetic field and a rotational electric field on particle charge. The second term in formula (\ref{SC ham gen}) describes interaction of particle charge with an external potential electric field. And the third term presents the action of an external magnetic field on the magnetic moments. The second group of terms consists of the two last terms. They describe the interparticle interaction. In this paper we consider the Coulomb and the spin-spin interactions presented by the fourth and fifth terms correspondingly.

Using explicit form of the Hamiltonian, we obtain the chain of equations, we truncate the chain of the QHD equations including the four equations only. These are the evolution equations for the particle concentration $n$, the velocity field $v^{\beta}$, the density of magnetic moment or spin $M^{\alpha}$, and the spin-current $J^{\alpha\beta}$.

The first step in derivation of the many-particle QHD equations is the definition of particle concentration. Which is the first collective quantum mechanical observable in our model. The particle concentration is the quantum mechanical average of the microscopic concentration
\begin{equation}\label{SC concentration}n(\textbf{r},t)=\int \Psi^{+}(R,t)\sum_{p}\delta(\textbf{r}-\textbf{r}_{p})\Psi(R,t) dR,\end{equation}
where we integrate over the 3N dimensional configurational space, and $dR=\prod_{p=1}^{N}d\textbf{r}_{p}$. The formula (\ref{SC concentration}) is more than the first collective quantum mechanical observable. Using of the particle concentration operator $\hat{n}=\sum_{p}\delta(\textbf{r}-\textbf{r}_{p})$ gives the projector of the 3N dimensional configurational space in the three dimensional physical space. Waves propagation, the charge-current flow, the spin-current flow happen in the three dimensional physical space. Consequently it is worthwhile to have a model, which explicitly describes the dynamic of quantum many-particle system in the physical space. The QHD is an example of such model.

We now differentiate the particle concentration with respect to time and find the continuity equation, where the particles current $\textbf{j}=n\textbf{v}$ emerges.

Hence, the first equation of the QHD set of equations is the continuity equation
\begin{equation}\label{SC continuity equation}\partial_{t}n+\nabla (n\textbf{v})=0.\end{equation}

At derivation of the continuity equation (\ref{SC continuity equation}) the explicit form of the particles current appears as
\begin{equation}\label{SC current}\textbf{j}=\int \sum_{p}\delta(\textbf{r}-\textbf{r}_{p})\frac{1}{2m_{p}}\biggl(\Psi^{+}(R,t)\textbf{D}_{p}\Psi(R,t)+h.c.\biggr) dR,\end{equation}
where h.c. means the hermitian conjugation.

Differentiating the function of current with respect to time, we obtain the momentum balance equations, this equation is an analog of the Euler equation
$$mn(\partial_{t}+\textbf{v}\nabla)v^{\alpha}+\partial_{\beta}p^{\alpha\beta}$$
$$-\frac{\hbar^{2}}{4m}\partial^{\alpha}\triangle
n+\frac{\hbar^{2}}{4m}\partial^{\beta}\Biggl(\frac{\partial^{\alpha}n\cdot\partial^{\beta}n}{n}\Biggr)$$
\begin{equation}\label{SC momentum balance eq}=enE^{\alpha}+\frac{e}{c}\varepsilon^{\alpha\beta\gamma}nv^{\beta}B^{\gamma}+M^{\beta}\nabla^{\alpha}B^{\beta},
\end{equation}
where $\textbf{E}$ and $\textbf{B}$ are the electric and magnetic fields, $\textbf{M}$ is the density of magnetic moments, $\varepsilon^{\alpha\beta\gamma}$ is the antisymmetric symbol (the Levi-Civita symbol), $p^{\alpha\beta}$ is the kinetic pressure tensor.
The momentum balance equation (\ref{SC momentum balance eq}) has usual form, we see that evolution of the velocity field caused by momentum current on thermal velocities $p^{\alpha\beta}$, the quantum Bohm potential specific for quantum kinematics (two terms in the left-hand side of equation (\ref{SC momentum balance eq}), which proportional to $\hbar^{2}$), and interaction, which is presented in the right-hand side of the momentum balance equation (\ref{SC momentum balance eq}). We derive this equation for charged spinning particles, thus the force field contains the density of the Lorentz force describing action of electromagnetic field on charges presented by  two first terms and force acting on the magnetic moment density from the magnetic field presented by the last term. We present the force field in the self-consistent field approximation. General form of the force field appearing in the Euler equation and introducing of the self-consistent field approximation are presented in Appendix A.

The second-fifth terms in the left-hand side of equation (\ref{SC momentum balance eq}) appear due to representation of the momentum flux $\Pi^{\alpha\beta}$. At derivation of the Euler equation the momentum flux $\Pi^{\alpha\beta}$ emerges in the following explicit form
$$\Pi^{\alpha\beta}=\int \sum_{p}\delta(\textbf{r}-\textbf{r}_{p})\frac{1}{2m_{p}}\times$$
\begin{equation}\times\biggl(\Psi^{*}(R,t)\hat{D}_{p}^{\beta}\hat{D}_{p}^{\alpha}\Psi(R,t)+h.c.\biggr) dR.\end{equation}
To represent the momentum flux via the hydrodynamic variables we need to introduce the velocity of a quantum particle $v_{i}^{\alpha}(R,t)$. It appears via the phase $S(R,t)$ of the many-particle wave function $\Psi(R,t)=a(R,t)e^{\imath S(R,t)}$. The velocity of i th particle is $v_{i}^{\alpha}(R,t)=\frac{\hbar}{m}\nabla_{i}S(R,t)$. We can also introduce the thermal velocity of i th particle $u_{i}^{\alpha}=v_{i}^{\alpha}(R,t)-v^{\alpha}(\textbf{r},t)$ as difference of the velocity of i th particle and the the center of mass velocity (the velocity field) $\textbf{v}(\textbf{r},t)=\textbf{j}(\textbf{r},t)/n(\textbf{r},t)$ (for details see Refs. \cite{MaksimovTMP 1999}, \cite{Andreev arxiv 11 3}).

The definition of magnetization
\begin{equation}\label{SC magnetization}M^{\alpha}(\textbf{r},t)=\int \sum_{p}\delta(\textbf{r}-\textbf{r}_{p})\Psi^{+}(R,t)\widehat{\sigma}^{\alpha}\Psi(R,t) dR,\end{equation}
appears at derivation of the Euler equation (\ref{SC momentum balance eq})

The equation of evolution of the magnetic moments
\begin{equation}\label{SC magn mom balance eq}\partial_{t}M^{\alpha}+\nabla^{\beta}J^{\alpha\beta}=\frac{2\gamma}{\hbar}\varepsilon^{\alpha\beta\gamma}M^{\beta}B^{\gamma}\end{equation}
is derived at differentiating of the magnetization with respect to time and using of the Schrodinger equation for the time derivatives of the wave function. This equation is a generalization of the Bloch equation. From equation (\ref{SC magn mom balance eq}) we see that evolution of magnetic moment density caused by both the spin current $J^{\alpha\beta}$ and interaction of the magnetic moments with the magnetic field. Charge of particles gives no interference in dynamics of the magnetic moment density.

The explicit form of spin-current in the many-particle system is
$$J^{\alpha\beta}=\int \sum_{p}\delta(\textbf{r}-\textbf{r}_{p})\frac{1}{2m_{p}}\times$$
\begin{equation}\label{SC spin current}\times\biggl(\Psi^{+}(R,t)\hat{D}_{p}^{\beta}\widehat{\sigma}_{p}^{\alpha}\Psi(R,t)+h.c.\biggr) dR.\end{equation}

Next equation is the equation of spin-current evolution. The
spin-current appears in the Bloch equation, and if we include the
spin-orbit and the spin-current interaction we get that the
spin-current gives contribution in the force field in the Euler
equation (\ref{SC momentum balance eq}), for example see Ref.
~\cite{Andreev IJMP 12}. The spin-current evolution equation for system of neutral particles is
$$\partial_{t}J^{\alpha\beta}+\partial^{\gamma}(J^{\alpha\beta}v^{\gamma})$$
\begin{equation}\label{SC magn mom current balance eq}=\frac{\gamma^{2}}{m}n\partial^{\beta}B^{\alpha}-\frac{2\gamma}{\hbar}\varepsilon^{\alpha\gamma\delta}B^{\gamma}J^{\delta\beta}.\end{equation}
for flux of the spin-current $J^{\alpha\beta\gamma}$ which
emerges in the second term in the left-hand side of equation
(\ref{SC magn mom current balance eq}). Its explicit form is
$$J^{\alpha\beta\gamma}=\int \sum_{p}\delta(\textbf{r}-\textbf{r}_{p})\frac{1}{4m_{p}^{2}}\times$$
\begin{equation}\label{SC spin current flux} \times\biggl(\Psi^{+}(R,t)\hat{D}_{p}^{\gamma}\hat{D}_{p}^{\beta}\widehat{\sigma}_{p}^{\alpha}\Psi(R,t)+h.c.\biggr) dR.\end{equation}
Definition of the flux of spin-current contains two operators of the long derivative $\hat{D}_{p}^{\alpha}$ when the spin-current contains one operator of the long derivative.
Since $J^{\alpha\beta\gamma}$ is the flux of the spin-current we use an approximate
formula $J^{\alpha\beta\gamma}=J^{\alpha\beta}v^{\gamma}$. In
general case $J^{\alpha\beta\gamma}$ has more complex structure
and contains additional contribution of both the thermal motion
and quantum kinematics as the quantum Bohm potential. The last one shows in the form of a term
analogous to the quantum Bohm potential.  We do not consider these
contributions and pay attention to the spin current evolution
caused by inter-particle interaction. We should pay special
attention to equation (\ref{SC magn mom current balance eq})
because all this paper is dedicated to consideration of the
influence of the spin-current evolution on dynamics of particles
system.  Equation (\ref{SC magn mom
current balance eq}) is presented for the chargeless spinning
particles, and this form will be used in the paper. However,
we now present it for the charged spinning particles
$$\partial_{t}J^{\alpha\beta}+\partial^{\gamma}(J^{\alpha\beta}v^{\gamma})=\frac{e}{m}M^{\alpha}E^{\beta}$$
\begin{equation}\label{SC magn mom current balance eq charge}+\frac{e}{mc}\gamma\varepsilon^{\beta\gamma\delta}J^{\alpha\gamma}B^{\delta}+\frac{\gamma^{2}}{m}n\partial^{\beta}B^{\alpha}-\frac{2\gamma}{\hbar}\varepsilon^{\alpha\gamma\delta}B^{\gamma}J^{\delta\beta}.\end{equation}

Equations (\ref{SC continuity equation})-(\ref{SC magn mom current balance eq}) take place for each species of particles. The electric $\textbf{E}$ and magnetic $\textbf{B}$ fields appearing in the equations (\ref{SC continuity equation})-(\ref{SC magn mom current balance eq}) are caused by charges, electric currents, and magnetic moments of medium and satisfy to the Maxwell equations. Thus the QHD equations for each species of particles connect by means of the Maxwell equations
$$\begin{array}{ccc} \nabla\textbf{B}=0 ,& \nabla\textbf{E}=4\pi\sum_{a}e_{a}n_{a} \end{array},$$
$$ \nabla\times\textbf{E}=0, $$
\begin{equation}\label{SC Maxwell eq}\nabla\times\textbf{B}=\frac{4\pi}{c}\sum_{a}e_{a}n_{a}\textbf{v}_{a}+4\pi \sum_{a}\nabla\times\textbf{M}_{a},\end{equation}
where subindex "a" describes the species of particles. The Maxwell equations (\ref{SC Maxwell eq}) presented here do not contain time derivatives of the electric and the magnetic fields, because we have derived the QHD equations from the non-relativistic theory.

\section{III. Dispersion equation}

We consider the collective eigen-waves in a system of the neutral spinning particles being in an external uniform magnetic field. Hence, we have deal with the paramagnetic and diamagnetic dielectrics. There are the two fundamental collective excitations in such physical systems, they are the sound waves and the spin waves. Following the QHD description of the three dimensional magnetized dielectrics ~\cite{Andreev PIERS 2012 Spin} one can show that there is one type of the spin waves in such systems. These waves have a constant eigen-frequency $\omega=2\gamma B_{0}/\hbar$, which is the cyclotron frequency, where $B_{0}$ is an external magnetic field. Here we have no dependency on wave vector, consequently the group velocity $\partial\omega/\partial k$ of these waves equal to zero.

We are interested in an interference of the spin-current evolution on the dispersion properties of the medium. To find the dispersion dependence of eigen-waves in the described systems we consider small amplitude excitations around an equilibrium state of the medium. We consider two different equilibrium states. In the first case we suppose that an equilibrium spin-current equals to zero, and in the second case we consider a medium with an equilibrium spin-current under condition that an equilibrium velocity field equals to zero.

For getting of solution we consider hydrodynamic variables as the sum of an equilibrium part and a small perturbation
$$\begin{array}{ccc}n=n_{0}+\delta n, &\textbf{E}=0+\textbf{E} \end{array},$$
$$\begin{array}{ccc}\textbf{B}=B_{0}\textbf{e}_{z}+\delta \textbf{B},&
\textbf{v}=0+\delta\textbf{v} \end{array},$$
$$\begin{array}{ccc}M^{\alpha}=M_{0}^{\alpha}+\delta M^{\alpha},& M_{0}^{\alpha}=\chi
B_{0}^{\alpha},& J^{\alpha\beta}=J^{\alpha\beta}_{0}+\delta J^{\alpha\beta},\end{array}$$
\begin{equation}\label{SC linearisation}\begin{array}{ccc}& p^{\alpha\beta}=p\delta^{\alpha\beta}, &
\delta p=mv_{F}^{2}\delta n_{a},
\end{array}\end{equation}
where $\delta^{\alpha\beta}$ is the Kronecker symbol, $v_{F}$ is the Fermi velocity,
$\chi_{a}=\kappa_{a}/\nu_{a}$ is the ratio between the equilibrium
magnetic susceptibility $\kappa_{a}$ and the magnetic permeability
$\nu_{a}=1+4\pi\kappa_{a}$. In the case $\kappa_{a}\ll 1$ we
have $\chi_{a}\simeq\kappa_{a}$. Substituting relations (\ref{SC linearisation})
in the set of equations (\ref{SC
continuity equation}), (\ref{SC momentum balance eq}), (\ref{SC
magn mom balance eq}) and (\ref{SC Maxwell eq}) and neglecting by the
nonlinear terms, we obtain a system of linear homogeneous
equations in the partial derivatives with constant coefficients.
Passing to the following representation for the small perturbations
$\delta f$
$$\delta f =f(\omega, k) exp(-\imath\omega t+\imath k x) $$
yields a homogeneous system of algebraic equations.

Even when we consider an equilibrium spin-current the linear set
of QHD equations splits on four independent sets. One of them
contains $n$, $\delta v_{x}$, $\delta M_{z}$, $J_{zx}$, the second
one contains $\delta M_{x}$, $\delta M_{y}$, $\delta J_{xx}$,
$\delta J_{yx}$, the third set includes $\delta J_{xy}$, $\delta
J_{yy}$, and fourth includes $\delta J_{xz}$, $\delta J_{yz}$. Two
last sets have the same solution, which is the cyclotron frequency
$\omega=2\gamma B_{0}/\hbar$. Hence, we have to consider solutions
of the first and second sets.

\subsection{A. The first group of dispersion branches}

The first set give the following dispersion equation
$$\omega^{4}+\omega^{2}\biggl(\frac{4\pi n_{0}\gamma^{2}}{m}-\upsilon^{2}\biggr)k^{2}$$
\begin{equation}\label{SC disp curr wave}+\frac{4\pi k^{3}M_{0}}{mn_{0}}J_{0}^{zx}\omega-\frac{4\pi n_{0}\gamma^{2}}{m}\upsilon^{2}k^{4}=0,\end{equation}
where
\begin{equation}\label{SC quantum vel}\upsilon^{2}=v_{F}^{2}+\frac{\hbar^{2}k^{2}}{4m^{2}},\end{equation}
and
\begin{equation}v_{F}=(3\pi^{2}n)^{1/3}\frac{\hbar}{m}.\end{equation}
We can see that a non-zero equilibrium spin-current leads to
existence of the additional term in the dispersion  equation.

In the absence of an equilibrium spin-current we have two solutions of the equation (\ref{SC disp curr wave})
\begin{equation}\label{SC new disp sol damp}\omega^{2}=-4\pi\gamma^{2}n_{0}k^{2}/m,\end{equation}
and
\begin{equation}\label{SC sound}\omega^{2}=\upsilon^{2}k^{2},\end{equation}
where solution (\ref{SC sound}) is the well-known sound wave. Dispersion of the sound wave consists of two parts (\ref{SC quantum vel}). The first of them is the usual linear term, which appears due to the Fermi pressure. The second term is the contribution of the quantum Bohm potential giving dispersion of the de Broglie wave. Moreover the evolution of spin-current leads to existence of the second solution (\ref{SC new disp sol damp}). This solution can be rewritten as
\begin{equation}\label{SC new disp sol damp lin fr}\omega=\pm\imath\sqrt{\frac{4\pi n_{0}}{m}}\gamma k.\end{equation}
It shows faster damping and gives no wave behavior.

The sound wave solution (\ref{SC sound}) and spin wave $\omega=2\gamma B_{0}/\hbar$ appear in the simpler model without account of the spin-current evolution. In this case a system of neutral spinning particles can be described by the continuity, Euler, and magnetic moment evolution equations, where we can put $J^{\alpha\beta}=M^{\alpha}v^{\beta}$ to close the set of equations. It has been mostly used for systems of charged spinning particles \cite{Andreev VestnMSU 2007}, \cite{Andreev IJMP 12}, \cite{Maksimov VestnMSU 2000}.

\begin{figure}
\includegraphics[width=8cm,angle=0]{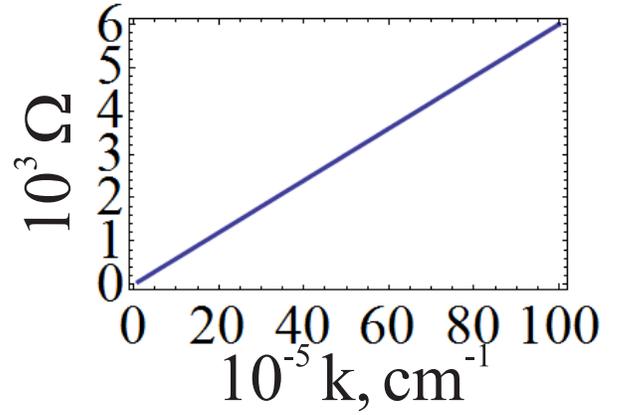}
\caption{\label{SC waves 1} (Color online) The figure describes
the dispersion dependence of the stable collective excitation described by equation (\ref{SC disp curr wave dimless}). This figure is obtained in the de-Broglie regime. We see that $\Omega$ approximately proportional to $k$. So this figure reveals quadratic dependence of the frequency $\omega$ of wave vector $k$.}
\end{figure}

\begin{figure}
\includegraphics[width=8cm,angle=0]{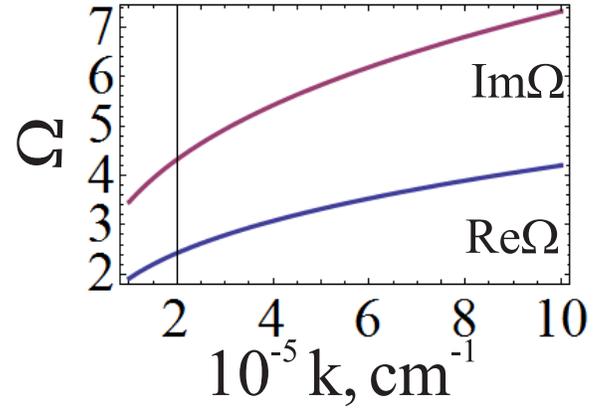}
\caption{\label{SC waves 2} (Color online) The figure describes the second solution of equation (\ref{SC disp curr wave dimless}) in the de-Broglie regime. This solution is unstable.)
The dispersion dependence and corresponding instability increment of the collective excitation are presented.}
\end{figure}

Presence of the equilibrium spin-current in the equation (\ref{SC disp curr wave}) makes it rather complicate. So, we are going to solve it numerically.
It seems reasonable to chouse $\Omega\equiv\omega/\lambda=\sqrt{m/(\pi n_{0})}\omega/(2\gamma k)$ as dimensionless frequency, where we introduced
\begin{equation}\label{SC lambda}\lambda^{2}=\frac{4\pi\gamma^{2}n_{0}}{m}k^{2}.\end{equation}
In this case we get equation (\ref{SC disp curr wave}) in the form of
\begin{equation}\label{SC disp curr wave dimless} \Omega^{4}+(1-\alpha)\Omega^{2}+\beta\Omega-\alpha=0,\end{equation}
where $\alpha=m\upsilon^{2}/(4\pi n_{0}\gamma^{2})$ is the parameter describing contribution of the quantum Bohm potential, $\beta=kM_{0}J_{0}^{zx}/(\gamma^{2}n_{0}^{2})$ is the parameter describing contribution of the equilibrium spin-current. On figures we present a limit case: de-Broglie regime when  $\alpha\simeq\alpha_{\hbar}=\hbar^{2}k^{2}/(16\pi mn_{0}\gamma^{2})$. The both parameters $\alpha$ and $\beta$ depend on module of the wave vector $k$. Consequently equation (\ref{SC disp curr wave dimless}) allows to get $\Omega(k)$ dependence. Equation (\ref{SC disp curr wave dimless}) gives two solutions. One of them is stable solution, which dispersion presented on Fig. (\ref{SC waves 1}). The second solution of equation (\ref{SC disp curr wave dimless}) shows an instability, which exists due to the existence of the equilibrium spin-current $J_{0}^{zx}$. This solution is presented on Fig. (\ref{SC waves 2}).

Solutions (\ref{SC new disp sol damp}) and (\ref{SC sound}) can be briefly written in terms of reduced frequency $\Omega$ and  $\alpha$, so we have $\Omega_{1}^{2}=-1$ and $\Omega_{2}^{2}=\alpha$.

Let us consider approximate solutions of equation (\ref{SC disp curr wave dimless}) at nonzero equilibrium spin-current under assumption that $\beta$ gives small contribution in dispersion of waves (\ref{SC sound}) and (\ref{SC new disp sol damp lin fr}). Thus solutions of equation (\ref{SC disp curr wave dimless}) emerge as
\begin{equation}\label{SC big disp first group sound}\omega=\upsilon k+\zeta_{1}\end{equation}
and
\begin{equation}\label{SC big disp first group damp}\omega=\pm\imath\sqrt{\frac{4\pi n_{0}}{m}}\gamma k+\zeta_{2},\end{equation}
where
$$\zeta_{i}=\frac{\beta\omega_{i}}{2\omega_{i}(2(\omega_{i}/\lambda)^{2}+1-\alpha)+\lambda\beta}$$
\begin{equation}\label{SC zeta} \simeq\frac{\beta}{2(2(\omega_{i}/\lambda)^{2}+1-\alpha)}.\end{equation}
$\omega_{i}$ is used for short representation of solutions (\ref{SC sound}) and (\ref{SC new disp sol damp lin fr}).
Including $\alpha\lambda^{2}=\upsilon k$ we come to the following dispersion relation for the sound wave
\begin{equation}\omega=\upsilon k+ \frac{\beta\lambda^{2}}{2(\upsilon^{2}k^{2}+\lambda^{2})}\equiv U_{1}^{*}k,\end{equation}
where we introduced a modified sound velocity $U_{1}^{*}$, which slightly depends on the wave vector $k$ via the quantum Bohm potential.
\begin{equation}\label{SC sound vel new} U_{1}^{*}=\upsilon+\frac{2\pi\gamma^{2}n_{0}(\beta/k)}{m\upsilon^{2}+4\pi\gamma^{2}n_{0}}.\end{equation}
Since $\beta >0$ and $\beta\sim k$ we see that $U_{1}^{*} > \upsilon$.

Contribution of the spin-current in unstable branch (\ref{SC big disp first group damp}) in considering limit
$$\omega=\pm\imath\lambda-\frac{\beta\lambda^{2}}{2(\upsilon^{2}k^{2}+\lambda^{2})}$$
\begin{equation}\label{SC big disp first group damp explicit}= \pm\imath\sqrt{\frac{4\pi n_{0}}{m}}\gamma k -\frac{2\pi\gamma^{2}n_{0}\beta}{m\upsilon^{2}+4\pi\gamma^{2}n_{0}}\end{equation}
appears to be real and negative. We see that real part of solution (\ref{SC big disp first group damp explicit}) is almost a linear function of the wave vector $k$, but it contains small additional dependence on the wave vector via $\upsilon$ containing contribution of the quantum Bohm potential. This results differs from previously considered case depicted on Fig.2. Formula (\ref{SC big disp first group damp explicit}) is obtained in the limit of small contribution of the equilibrium spin-current. When Fig.2 is obtained for a finite value of the equilibrium spin-current, so it reveals an interesting dependence of real and imaginary parts of the frequency on the wave vector discussed above.

\subsection{B. Wave dispersion for the second group of variables}

The second set of equations containing evolution of $\delta
M_{x}$, $\delta M_{y}$, $\delta J_{xx}$, $\delta J_{yx}$ gives the
following dispersion equation
$$\omega^{2}+8\pi\frac{\Omega_{\gamma}\gamma}{\hbar\omega}k J^{zx}_{0}$$
\begin{equation}\label{SC disp eq for MMJJ}+4\pi\frac{k}{\omega}(\omega^{2}+\Omega^{2}_{\gamma})\frac{\omega k n_{0}\gamma^{2}\hbar+\Omega_{\gamma}2\gamma m J^{zx}_{0}}{m\hbar(\omega^{2}-\Omega^{2}_{\gamma})}-\Omega_{\gamma}^{2}(1-4\pi\xi)=0,\end{equation}
where $\Omega_{\gamma}=\frac{2\gamma B_{0}}{\hbar}$ and $\xi=M_{0}/B_{0}$. For the ferromagnetic samples $\xi$ is larger than one $\xi\gg 1$, while $\xi$ is rather small $\xi\ll 1$ for the para- and the dia-magnetics.

In the absence of the equilibrium spin-current this equation is simplified to
\begin{equation}\label{SC disp eq for MMJJ simp}\omega^{4}+\biggl(\lambda^{2}-2\Omega^{2}_{\gamma}(1-2\pi\xi)\biggr)\omega^{2}+\Omega^{2}_{\gamma}\biggl(\lambda^{2}+\Omega^{2}_{\gamma}(1-4\pi\xi)\biggr)=0.\end{equation}
Solving this equation we get
$$\omega^{2}=\Omega^{2}_{\gamma}(1-2\pi\xi)-\frac{1}{2}\lambda^{2}$$
\begin{equation}\label{SC spin new waves gen} \pm\sqrt{\frac{1}{4}\lambda^{4}-2\lambda^{2}\Omega^{2}_{\gamma}(1-\pi\xi)+16\pi^{2}\xi^{2}\Omega^{4}_{\gamma}}.\end{equation}
These are new solutions. Their occurrence is caused by the account of the spin-current evolution, so we called them spin-current waves.

In the small $k$ limit we come to
\begin{equation}\label{SC spin new waves small k} \omega^{2}=\Omega^{2}_{\gamma}\left(\begin{array}{ccc}1+2\pi\xi & &\\ 1-6\pi\xi &  &\\\end{array}\right)+\lambda^{2}\left(\begin{array}{ccc} -\frac{1}{4}-\frac{1}{4\pi\xi} & &\\ -\frac{3}{4}+\frac{1}{4\pi\xi} &  &\\\end{array}\right).\end{equation}
In the paramagnetic limit formula (\ref{SC disp eq for MMJJ simp}) gives
\begin{equation}\label{SC spin new waves param lim} \omega_{\xi}^{2}=\Omega^{2}_{\gamma}-\frac{1}{2}\lambda^{2}\pm\lambda\sqrt{\frac{1}{4}\lambda^{2}-2\Omega^{2}_{\gamma}}.\end{equation}
At large $k$ at small magnetic field ($\lambda^{2}\gg \Omega^{2}_{\gamma}$) only one solution exists, which corresponds to the sign plus in front of the square root in formula (\ref{SC spin new waves gen}). This solution appears as
\begin{equation}\label{SC } \omega_{\lambda, +}=\Omega_{\gamma}\sqrt{16\pi^{2}\xi^{2}\frac{\Omega^{2}_{\gamma}}{\lambda^{2}}-1},\end{equation}
since $\frac{\Omega^{2}_{\gamma}}{\lambda^{2}}\ll 1$ we have that
this solution exists at $\xi >\frac{\lambda}{4\pi\mid\Omega_{\gamma}\mid}\gg 1$. It corresponds to large magnetization $M_{0}$.

Equation (\ref{SC disp eq for MMJJ}) can be rewritten as an equation of fifth degree. One of its solution is $\omega=0$. So we have equation of fourth degree
$$\omega^{4}+\biggl(\lambda^{2}-2\Omega_{\gamma}^{2}(1-2\pi\xi)\biggr)\omega^{2}$$
\begin{equation}\label{SC disp eq for MMJJ reduced}+\vartheta\omega+\Omega_{\gamma}^{2}\biggl(\Omega_{\gamma}^{2}(1-4\pi\xi)+\lambda^{2}\biggr)=0,\end{equation}
where
\begin{equation}\vartheta=16\pi\gamma\Omega_{\gamma}J_{0}^{zx}k/\hbar.\end{equation}
This equation (\ref{SC disp eq for MMJJ reduced}) differs from (\ref{SC disp eq for MMJJ simp}) by one term only. It is the third term $\vartheta\omega$. Thus we have changing of solutions (\ref{SC spin new waves gen}) caused by the equilibrium spin-current $\vartheta\sim J_{0}^{zx}$. Solutions of equation (\ref{SC disp eq for MMJJ reduced}) we calculated approximately under assumption that the equilibrium spin-current gives small contribution in dispersion dependence. Designating solutions of equation (\ref{SC disp eq for MMJJ simp}) presented by formula (\ref{SC spin new waves gen}) as $\varpi$. Then solutions of equation (\ref{SC disp eq for MMJJ reduced}) appear as
$$\omega=\varpi+\frac{\vartheta\varpi}{2\varpi(2\varpi^{2}+\lambda^{2}-2\Omega_{\gamma}^{2}(1-2\pi\xi))+\vartheta}$$
\begin{equation}\label{SC disp eq for MMJJ appr fprm 1}\simeq\varpi+\frac{\vartheta}{2(2\varpi^{2}+\lambda^{2}-2\Omega_{\gamma}^{2}(1-2\pi\xi))}.\end{equation}
Using explicit form of $\varpi$ in the second term of formula (\ref{SC disp eq for MMJJ appr fprm 1}) we come to the following formula
\begin{equation}\label{SC disp eq for MMJJ appr fprm 2}\omega=\varpi\pm\frac{\vartheta}{2\sqrt{\lambda^{4}-8\lambda^{2}\Omega_{\gamma}^{2}(1-\pi\xi)+64\pi^{2}\xi^{2}\Omega_{\gamma}^{4}}}.\end{equation}
Signs plus and minus in front of the second term of formula (\ref{SC disp eq for MMJJ appr fprm 2}) correspond to signs in formula (\ref{SC spin new waves gen}).

\subsection{C. Discussion}

Formulas (\ref{SC new disp sol damp}), (\ref{SC sound}), (\ref{SC
spin new waves gen}) and the cyclotron frequency $\omega=2\gamma
B_{0}/\hbar$ appear as solutions of a dispersion equation obtained
at account of the spin-current evolution in the absence of the
equilibrium spin-current.

Solution (\ref{SC new disp sol damp}) reveals two branches
(\ref{SC new disp sol damp lin fr}). One of them has an increasing
amplitude. Another one has a decreasing amplitude. Since we have
no source of the energy in the system we have no mechanism for the
amplitude increasing. So, we conclude that the decreasing branch
takes place in considering case. This decreasing solution gives no
contribution in the spectrum as it reveals monotonic decreasing of the amplitude (non oscillating solution). Thus we have got left the two unchanged solutions: the sound
wave (\ref{SC sound}) and the spin wave $\omega=2\gamma
B_{0}/\hbar$, which can be found with no account of the
spin-current evolution. We have also found solutions (\ref{SC spin
new waves gen}), which present the two spin-current waves reaching the wave
spectrum of magnetized dielectrics.

At this step we can conclude that the account of the spin-current evolution, without changing of conditions system being at, we obtained additional information about processes happen in the system. In considering case we have got the two additional wave branches (the spin-current waves) described by formula (\ref{SC spin new waves gen}). Formula (\ref{SC spin new waves small k}) reveals that at small $k$ and large enough external magnetic field $B_{0}$ the cyclotron frequency $\Omega_{\gamma}=2\gamma B_{0}/\hbar$ gives main contribution in the dispersion of the spin-current waves.

Consideration of the spin-current as an independent physical variable gives us possibility to consider some conditions, which can not be included in more simple model. One of these conditions is the existence of an equilibrium spin-current with the zero equilibrium velocity field $v_{0}=0$. Presence of an equilibrium spin-current leads to changes of solutions (\ref{SC new disp sol damp}) and (\ref{SC sound}), and to complication of dispersion equation (\ref{SC disp eq for MMJJ simp}) (degree of this equation increases, so we get equation (\ref{SC disp eq for MMJJ})). However it gives no change in the dispersion of spin waves with the cyclotron frequency $\omega=2\gamma B_{0}/\hbar$.

Presence of an equilibrium spin-current $J_{0}^{zx}$ leads to change of the solutions (\ref{SC sound}) and (\ref{SC new disp sol damp lin fr}). Real part of solution (\ref{SC new disp sol damp lin fr}) appears due to $J_{0}^{zx}$. In the de-Brolie regime it is pictured by lowest curve on Fig. (2). It reveals as a curve with two linear areas, one at small $k$, and the second one at $k\geq 4\cdot 10^{5}$ cm$^{-1}$. They connect smoothly around $k=3\cdot 10^{5}$ cm$^{-1}$. Thus, we can assume $\Omega=\nu_{i}k$, with different $\nu_{i}$ for each area. Consequently, we have $Re\omega=\sqrt{\frac{\pi n_{0}}{m}}2\gamma\nu_{i}k^{2}$. Imaginary part of solution $Im\omega$ is pictured by upper curve on Fig.(2). Its form is similar to the form of the real part of the frequency $Re\omega$, but it has larger value. Limit of small contribution of the equilibrium spin-current allows to obtain some analytical solution (\ref{SC big disp first group damp explicit}). In this approximation the equilibrium spin-current $J_{0}^{zx}$ gives rise to appearance of the negative real part of the frequency. It gives no changes in the imaginary part of the solution.

Let us discuss the sound wave. In the absence of an equilibrium spin-current it is described by formula (\ref{SC sound}). In the presence of an equilibrium spin-current we numerically consider a limit case: the de-Broglie regime. In the de-Broglie regime we find that the equilibrium spin-current does not change form of the dispersion curve (formula (\ref{SC sound}) gives us $\omega=\frac{\hbar k^{2}}{2m}$ in the de-Broglie regime). Thus we see that $\Omega$ linearly depends on the wave vector $k$, and $\omega\sim k^{2}$. The approximation of small equilibrium spin-current lets to trace contribution of the equilibrium spin-current on the sound wave analytically. In the de-Broglie regime an addition to spectrum of the free quantum particles as
$$\triangle\omega_{S}=\frac{8\pi mn_{0}\gamma^{2}\beta}{\hbar^{2}k^{2}+16\pi mn_{0}\gamma^{2}}\sim\frac{k}{k^{2}+\chi^{2}},$$
where $\chi$ does not depend on the wave vector $k$. One can see that the frequency shift is positive and decreases at the increasing of the wave vector. In the Fermi regime we find that the sound velocity increases on a constant value defined by formula (\ref{SC sound vel new}).

Let us discuss now contribution of the equilibrium spin-current in the dispersion of the spin-current waves. Formula (\ref{SC disp eq for MMJJ appr fprm 2}) shows that the spin-current wave having plus (minus) in front of the square root in formula (\ref{SC spin new waves gen}) gets positive (negative) contribution of the equilibrium spin-current in the dispersion dependence. So we find an increasing (a decreasing) of the frequency. Including the fact that $\vartheta\sim k$ and $\lambda\sim k$ we have the following dependence of the last term in formula (\ref{SC disp eq for MMJJ appr fprm 2}) (a frequency shift caused by the equilibrium spin-current) on the wave vector appears as
$$\triangle\omega_{J}=\frac{8\sqrt{\pi m}\Omega_{\gamma}J_{0}^{zx}}{\hbar\sqrt{n_{0}}}\frac{k}{\sqrt{k^{4}-\tilde{a}k^{2}+\tilde{b}}},$$
where $\tilde{a}=2m(1-\pi\xi)\Omega_{\gamma}^{2}/(n_{0}\gamma^{2})$ and $\tilde{b}=4m^{2}\xi^{2}\Omega_{\gamma}^{4}/(n_{0}^{2}\gamma^{4})$ are positive constants. At $k\rightarrow 0$ we obtain $\triangle\omega_{J}\rightarrow 0$. $\triangle\omega_{J}$ increases with increasing of the wave vector. However $\triangle\omega_{J}$ reaches its maximum at an intermediate wave vector $k_{0}=\frac{\sqrt{2m\xi}\Omega_{\gamma}}{\gamma\sqrt{n_{0}}}$. This maximum value of the frequency shift is
$$\triangle\omega_{J}(k_{0})=\frac{8\sqrt{\pi m}\Omega_{\gamma}J_{0}^{zx}}{\hbar\sqrt{n_{0}}}\frac{1}{\sqrt{2\sqrt{\tilde{b}}-\tilde{a}}}.$$
At large $k$ the frequency shift $\triangle\omega_{J}$ decreases as $1/k$.

\section{IV. Conclusion}

To get influence of the spin-current on dynamics of magnetized
dielectrics we have derived equation of the spin-current
evolution as a part of the set of the QHD equations. In the result we have set of four equations: the continuity equation (particle number evolution equation), the Euler equation (momentum balance equation), the Bloch equation (magnetic moment balance equation), and equation of the spin-current evolution. These equations are used in the self-consistent field approximation.

With no account of the spin-current evolution equation we find two wave solutions: the sound wave and one spin wave solution.  Including the spin-current evolution equation leads to account of new solutions. We found three new wave solutions. Two of them have frequencies near the cyclotron frequency. These solutions make spectrum of spin waves richer. It appears as a splitting of one spin wave branch on three branches. The third solution has negative square of the frequency and reveal monotonic damping of perturbation amplitude. These solutions emerge when the equilibrium spin-current equals to zero. Account of an equilibrium spin-current does not give new solution, but a change of wave dispersion was obtained.

\section{Appendix A: General form of the force field and the self-consistent field approximation}

In the Euler equation (\ref{SC momentum balance eq}) the force
field
\begin{equation}\label{SC Force SCF 1} F^{\alpha}=enE^{\alpha}+\frac{e}{c}\varepsilon^{\alpha\beta\gamma}nv^{\beta}B^{\gamma}+M^{\beta}\nabla^{\alpha}B^{\beta} \end{equation}
is presented in the self-consistent field approximation. Here we
are going to present general form of this force field and explain
how we got formula (\ref{SC Force SCF 1}) from the general
formula.

Force field consists of two parts
\begin{equation}\label{SC Force sum} F^{\alpha}=F^{\alpha}_{ext}+F^{\alpha}_{int}. \end{equation}
The first of them is the force of the particle interaction with an external field
\begin{equation}\label{SC Force ext} F^{\alpha}_{ext}=enE^{\alpha}_{ext}+\frac{e}{c}\varepsilon^{\alpha\beta\gamma}nv^{\beta}B^{\gamma}_{ext}+M^{\beta}\nabla^{\alpha}B^{\beta}_{ext}, \end{equation}
and the second part is the inter-particle interactions
$$F^{\alpha}_{int}=-e^{2}\int (\nabla^{\alpha} G(\textbf{r}-\textbf{r}')) n_{2}(\textbf{r},\textbf{r}',t) d\textbf{r}' $$
\begin{equation}\label{SC Force int}
+\int (\nabla^{\alpha} G^{\beta\gamma}(\textbf{r}-\textbf{r}')) M_{2}^{\beta\gamma}(\textbf{r},\textbf{r}',t) d\textbf{r}', \end{equation}
where
\begin{equation}\label{SC two part conc} n_{2}(\textbf{r},\textbf{r}',t)=\int \sum_{p,n\neq p} \delta(\textbf{r}-\textbf{r}_{p})\delta(\textbf{r}-\textbf{r}'_{n}) \Psi^{*}(R,t)\Psi(R,t) dR\end{equation}
is the two-particle concentration, and
$$M_{2}^{\alpha\beta}(\textbf{r},\textbf{r}',t)=\int \sum_{p,n\neq p} \delta(\textbf{r}-\textbf{r}_{p})\delta(\textbf{r}-\textbf{r}'_{n})\times$$
\begin{equation}\label{SC two part magn}  \times\mu_{B}^{2}\Psi^{*}(R,t)\sigma^{\alpha}_{p}\sigma^{\beta}_{n}\Psi(R,t) dR\end{equation}
is the two-particle magnetization.

It has been shown that a two-particle function
$f_{2}(\textbf{r},\textbf{r}',t)$ (see formulas (\ref{SC two part
conc}) and (\ref{SC two part magn})) appears as a sum of two terms
$f_{2}(\textbf{r},\textbf{r}',t)=f(\textbf{r},t)f(\textbf{r}',t)+g_{2}(\textbf{r},\textbf{r}',t)$.
The first of the two terms is the product of corresponding
one-particle functions. It corresponds to the self-consistent
field approach suitable for the long-range interaction. The second
term gives contribution of quantum correlations, particularly the
exchange correlation.

Thus, in the self-consistent field approximation we have that the
two-particle concentration represents as the product of the
concentrations in points $\textbf{r}$ and $\textbf{r}'$:
$n_{2}(\textbf{r},\textbf{r}',t)=n(\textbf{r},t)n(\textbf{r}',t)$;
and the two-particle magnetization gives us the product of the
one-particle magnetization
$M^{\alpha\beta}_{2}(\textbf{r},\textbf{r}',t)=M^{\alpha}(\textbf{r},t)M^{\beta}(\textbf{r}',t)$.
Putting these representations in formula (\ref{SC Force int}) we
can introduce electric field caused by the electric charges and
the magnetic field caused by magnetic moments. These fields emerge
as
\begin{equation}E^{\alpha}=-e\nabla^{\alpha}\int  G(\textbf{r}-\textbf{r}') n(\textbf{r}',t) d\textbf{r}',\end{equation}
and
\begin{equation}B^{\alpha}=\int  G^{\alpha\beta}(\textbf{r}-\textbf{r}') M^{\beta}(\textbf{r}',t) d\textbf{r}'.\end{equation}
They satisfy the Maxwell equations (\ref{SC Maxwell eq}).  Using
these fields we come to the force field (\ref{SC Force SCF 1}).

\end{document}